\documentclass[a4paper]{JACoW-GSI}
\usepackage{graphicx}
\usepackage{url}

\begin{document}
\title{Recent experimental results on charmed baryons}

\author[1]{M. J. Charles\thanks{m.charles1@physics.ox.ac.uk}}
\affil[1]{University of Oxford, Oxford, OX1 3RH, UK}

\maketitle

\section{Introduction}

This paper summarizes recent experimental results on charmed baryons,
as presented at the 
  Third International Workshop on Charm Physics,
  20--22 May 2009,
  Heidelberg, Germany~\cite{bib:charm2009}.
Two main topics are covered: first, the spectroscopy of charmed
baryon states; second, the production of charmed baryons in
the decays of $B$ mesons and from the $e^+ e^-$ continuum.

\section{Spectroscopy}

There has been a steady progression of results on charmed baryon
spectroscopy from the $e^+ e^-$ $B$-factories BABAR~\cite{bib:babarDetector}
and Belle~\cite{bib:belleDetector} in
recent years, building on the foundations laid by
CLEO, FOCUS, E687, ARGUS, and others~\cite{bib:pdg}.
We summarize new and interesting results for
$\Sigma_c$, $\Xi_c$, and then $\Omega_c$ states.

\subsection{$\Sigma_c$ states}

\begin{figure}[htb]
  \centering
  \includegraphics*[width=80mm]{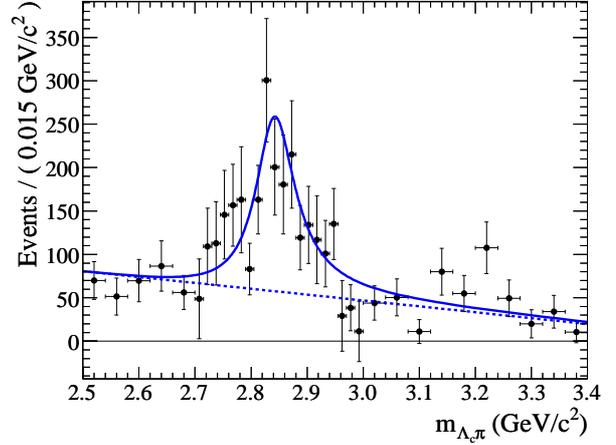}
  \caption{The $m(\Lambda_c^+ \pi^-)$ invariant mass distribution from
    $B \to \Lambda_c^+ \bar{p} \pi^-$ data at BABAR, using a likelihood
    weighting technique to suppress background. From~\cite{bib:babarStephanie}.
  }
  \label{fig:Sc2800}
\end{figure}

BABAR has recently published a study of the decay $B^- \to \Lambda_c^+ \bar{p} \pi^-$
in a sample of $383 \times 10^6$ $\Upsilon(4S) \to B \overline{B}$ decays~\cite{bib:babarStephanie,bib:disclaimer1}
measuring
$\mathcal{B}(B^- \to \Lambda_c^+ \bar{p} \pi^-) = (3.38 \pm 0.12 \pm 0.12 \pm 0.88) \times 10^{-4}$~\cite{bib:disclaimer2}.
This value is somewhat larger than the previous Belle result of
$(2.01 \pm 0.15 \pm 0.20 \pm 0.52) \times 10^{-4}$
from $152 \times 10^6$ $\Upsilon(4S) \to B \overline{B}$ decays~\cite{bib:belleStephanie}.
Both papers see clear signals for $B^- \to \Sigma_c(2455)^0 \bar{p}$ in the Dalitz plot, but a small or
no signal for $B^- \to \Sigma_c(2520)^0 \bar{p}$. 
Both papers also observed a $\Lambda_c^+ \bar{p}$ threshold enhancement, discussed later.
In addition, BABAR noted a broad, statistically
significant ($5.8\sigma$) peak in $\Lambda_c^+ \pi^-$
which is well-described by a relativistic $D$-wave Breit-Wigner with mass $2846 \pm 8$~MeV$/c^2$
and width $86^{+33}_{-22}$~MeV, as illustrated in Fig.~\ref{fig:Sc2800}.
This region of the Dalitz plot is populated in the Belle
data, but it is not clear whether a significant structure with the same parameters is present.
The closest known resonance is the $\Sigma_c(2800)^0$~\cite{bib:belleSc2800}:
the widths are consistent but the mass obtained by BABAR differs from the nominal value,
$2802 \pm 4 \pm 7$~MeV$/c^2$, by around three standard deviations. This difference is intriguing but
not conclusive. An updated study by Belle with their large $\Upsilon(4S)$ dataset could
settle this question and establish whether BABAR saw the $\Sigma_c(2800)^0$ with a downward
fluctuation in mass or an entirely new state.

\begin{figure}[htb]
  \centering
  \includegraphics*[width=80mm]{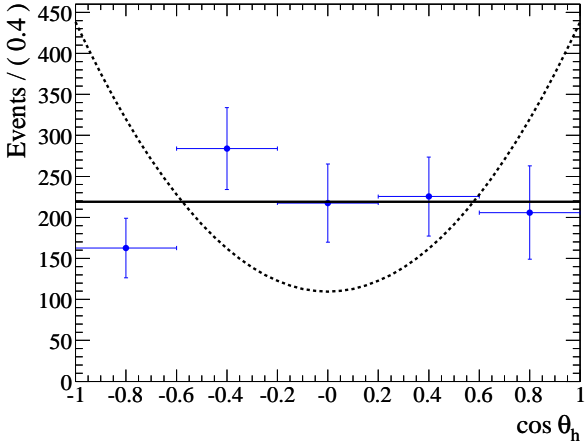}
  \caption{The helicity angle distribution for $\Sigma_c(2455)^0$ candidates in
    $B \to \Lambda_c^+ \bar{p} \pi^-$ data at BABAR.
    From~\cite{bib:babarStephanie}.
  }
  \label{fig:Sc2455}
\end{figure}

BABAR also used the large $B^- \to \Sigma_c(2455)^0 \bar{p}$ signal to determine the
spin of the $\Sigma_c(2455)^0$; the state is predicted to have $J^P = 1/2^+$
in the quark model but neither quantum number has been measured.
BABAR performed an angular analysis of the decay under the assumption that
$J(\Lambda_c^+) = 1/2$, testing the data for consistency with a spin of 1/2
or 3/2 for the $\Sigma_c(2455)^0$. Defining the helicity angle $\theta_h$
as the angle between the momentum vector of the $\Lambda_c^+$ and the
momentum vector of the recoiling $B$-daughter $\bar{p}$ in the rest frame
of the $\Sigma_c(2455)^0$, the expected angular distributions are:
\begin{eqnarray*}
  J(\Sigma_c(2455)^0) = 1/2: & \frac{\mathrm{d}N}{\mathrm{d} \cos \theta_h} & \propto 1 \\
  J(\Sigma_c(2455)^0) = 3/2: & \frac{\mathrm{d}N}{\mathrm{d} \cos \theta_h} & \propto 1 + 3 \cos^2 \theta_h
  .
\end{eqnarray*}
The measured helicity angle distribution,
shown in Fig.~\ref{fig:Sc2455}, is found to be consistent with spin-1/2 but
inconsistent with spin-3/2 by more than four standard deviations.

\subsection{$\Xi_c$ states}

There has been a flurry of results on new $\Xi_c$ states in the past three years,
beginning with the first observations of the $\Xi_c(2980)$ and $\Xi_c(3077)$ isodoublets
by Belle in the $\Lambda_c^+ K^- \pi^+$ and $\Lambda_c \overline{K}^0 \pi^+$ final
states with 462~fb$^{-1}$ of data~\cite{bib:belleXc2980}. These were subsequently confirmed
BABAR~\cite{bib:babarXc2980}, who also reported two other possible structures,
denoted $\Xi_c(3055)^+$ and $\Xi_c(3123)^+$, in 384~fb$^{-1}$ of data.
Belle has also observed the $\Xi_c(2980)$
in the final states $\Xi_c(2645)^0 \pi^+$ and $\Xi_c(2645)^+ \pi^-$
in a separate study of 414~fb$^{-1}$ of data~\cite{bib:belleXc2980b}.
The current understanding of these states is summarized below.
The resonance parameters are given in Table~\ref{tab:Xc2980}.

\begin{table*}
\centering
\caption{
  Results on new $\Xi_c$ resonances. The mass, width, and (where given) statistical
  significance are listed.
}
\begin{tabular}{llllr}
  \hline
  Source & \multicolumn{1}{c}{Resonance and final state} & Mass (MeV$/c^2)$ & Width (MeV) & \multicolumn{1}{c}{Signif.} \\ 
  \hline
  BABAR~\cite{bib:babarB2LcLcK} & $\Xi_c(2930)^0 \to \Lambda_c^+ K^-$ & $2931 \pm 3 \pm 5$ & $36 \pm 7 \pm 1$ & \\
  \hline
  Belle~\cite{bib:belleXc2980} & $\Xi_c(2980)^+ \to \Lambda_c^+ K^- \pi^+$ & $2978.5 \pm 2.1 \pm 2.0$ & $43.5 \pm 7.5 \pm 7.0$ & $6.3\sigma$ \\
  BABAR~\cite{bib:babarXc2980} & $\Xi_c(2980)^+ \to \Lambda_c^+ K^- \pi^+$ & $2969.3 \pm 2.2 \pm 1.7$ & $27 \pm 8 \pm 2$ & $>9.0\sigma$ \\
  Belle~\cite{bib:belleXc2980b}& $\Xi_c(2980)^+ \to \Xi_c(2645)^0 \pi^+$ & $2967.7 \pm 2.3 ^{+1.1}_{-1.2}$ & $18 \pm 6 \pm 3$ & $7.3\sigma$ \\
  \hline
  Belle~\cite{bib:belleXc2980} & $\Xi_c(2980)^0 \to \Lambda_c^+ \overline{K}^0 \pi^-$ & $2977.1 \pm 8.8 \pm 3.5$ & 43.5 (fixed) & $2.0\sigma$ \\
  BABAR~\cite{bib:babarXc2980} & $\Xi_c(2980)^0 \to \Lambda_c^+ \overline{K}^0 \pi^-$ & $2972.9 \pm 4.4 \pm 1.6$ & $31 \pm 7 \pm 8$ & $1.7\sigma$ \\
  Belle~\cite{bib:belleXc2980b}& $\Xi_c(2980)^0 \to \Xi_c(2645)^+ \pi^-$ & $2965.7 \pm 2.4 ^{+1.1}_{-1.2}$ & $15 \pm 6 \pm 3$ & $6.1\sigma$ \\
  \hline
  BABAR~\cite{bib:babarXc2980} & $\Xi_c(3055)^+ \to \Sigma_c(2455)^+ K^-$ & $3054.2 \pm 1.2 \pm 0.5$ & $17 \pm 6 \pm 11$ & $6.4\sigma$ \\
  \hline
  Belle~\cite{bib:belleXc2980} & $\Xi_c(3077)^+ \to \Lambda_c^+ K^- \pi^+$ & $3076.7 \pm 0.9 \pm 0.5$ & $6.2 \pm 1.2 \pm 0.8$ & $9.7\sigma$ \\
  BABAR~\cite{bib:babarXc2980} & $\Xi_c(3077)^+ \to \Lambda_c^+ K^- \pi^+$ & $3077.0 \pm 0.4 \pm 0.2$ & $5.5 \pm 1.3 \pm 0.6$ & $>9.0\sigma$ \\
  \hline
  Belle~\cite{bib:belleXc2980} & $\Xi_c(3077)^0 \to \Lambda_c^+ \overline{K}^0 \pi^-$ & $3082.8 \pm 1.8 \pm 1.5$ & $5.2 \pm 3.1 \pm 1.8$ & $5.1\sigma$ \\
  BABAR~\cite{bib:babarXc2980} & $\Xi_c(3077)^0 \to \Lambda_c^+ \overline{K}^0 \pi^-$ & $3079.3 \pm 1.1 \pm 0.2$ & $5.9 \pm 2.3 \pm 1.5$ & $4.5\sigma$ \\
  \hline
  BABAR~\cite{bib:babarXc2980} & $\Xi_c(3123)^+ \to \Sigma_c(2520)^+ K^-$ & $3122.9 \pm 1.3 \pm 0.3$ & $4.4 \pm 3.4 \pm 1.7$ & $3.0\sigma$ \\
  \hline
\end{tabular}
\label{tab:Xc2980}
\end{table*}

\begin{figure}[htb]
  \centering
  \begin{minipage}[b]{40mm}
    \centering
    \setlength{\unitlength}{0.47mm}
    \begin{picture}(95,85)
      \put(65,77){\bf (a)}
      \put(-5,35){\rotatebox{90}{\footnotesize Events / (5~{\rm MeV}/c$^2$)}}
      \put(6,-1){{\footnotesize $m(\Xi_c(2645)^0\pi^+)$ [GeV/c$^2$]}}
      \includegraphics[height=4.47cm,width=4.0cm]{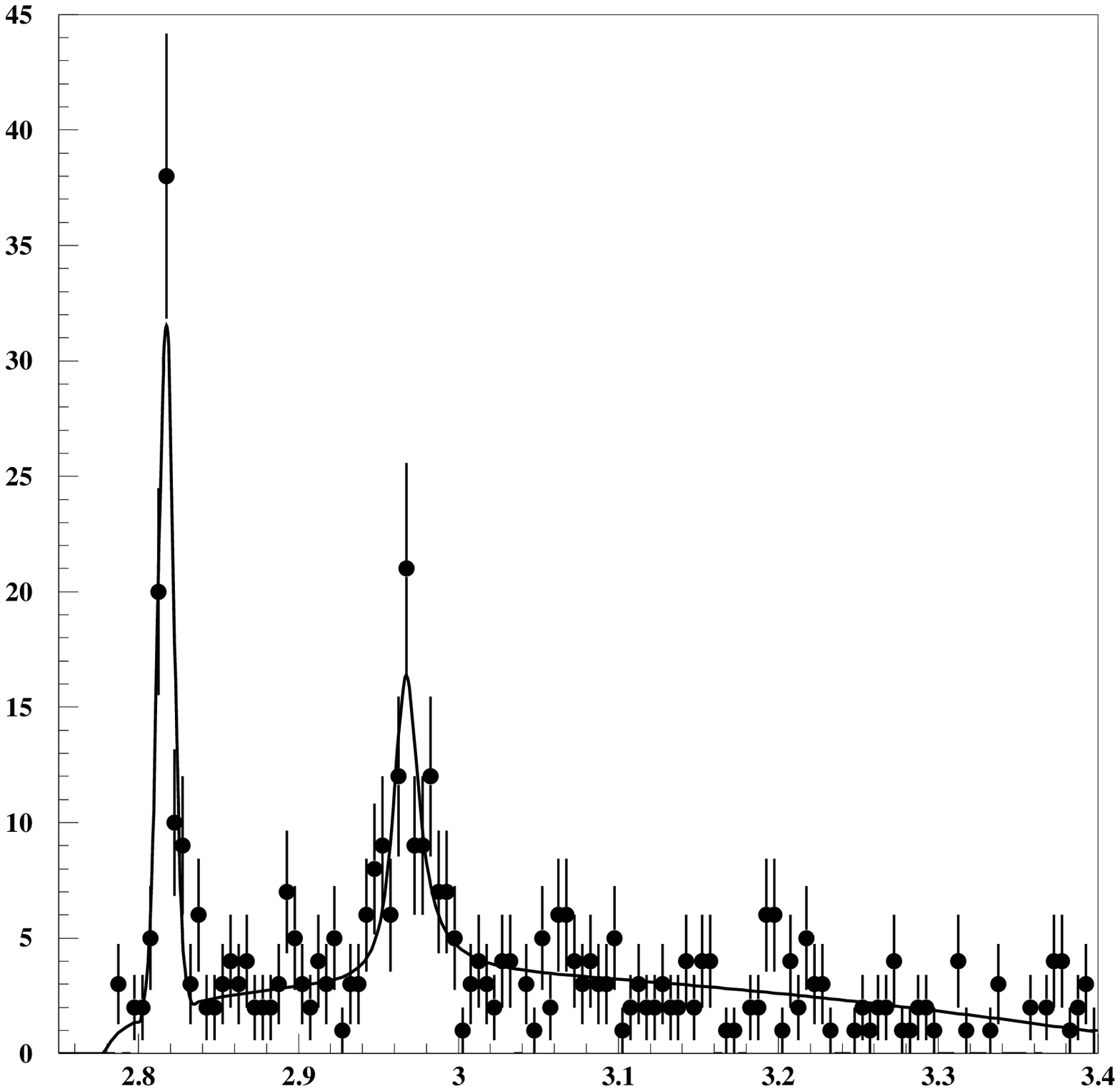}
    \end{picture}
  \end{minipage}\hfill
  \begin{minipage}[b]{40mm}
    \centering
    \setlength{\unitlength}{0.47mm}
    \begin{picture}(95,85)
      \put(65,77){\bf (b)}
      \put(5,-1){{\footnotesize $m(\Xi_c(2645)^+\pi^-)$ [GeV/c$^2$]}}
      \put(-5,35){\rotatebox{90}{\footnotesize Events / (5~{\rm MeV}/c$^2$)}}
      \includegraphics[height=4.47cm,width=4.0cm]{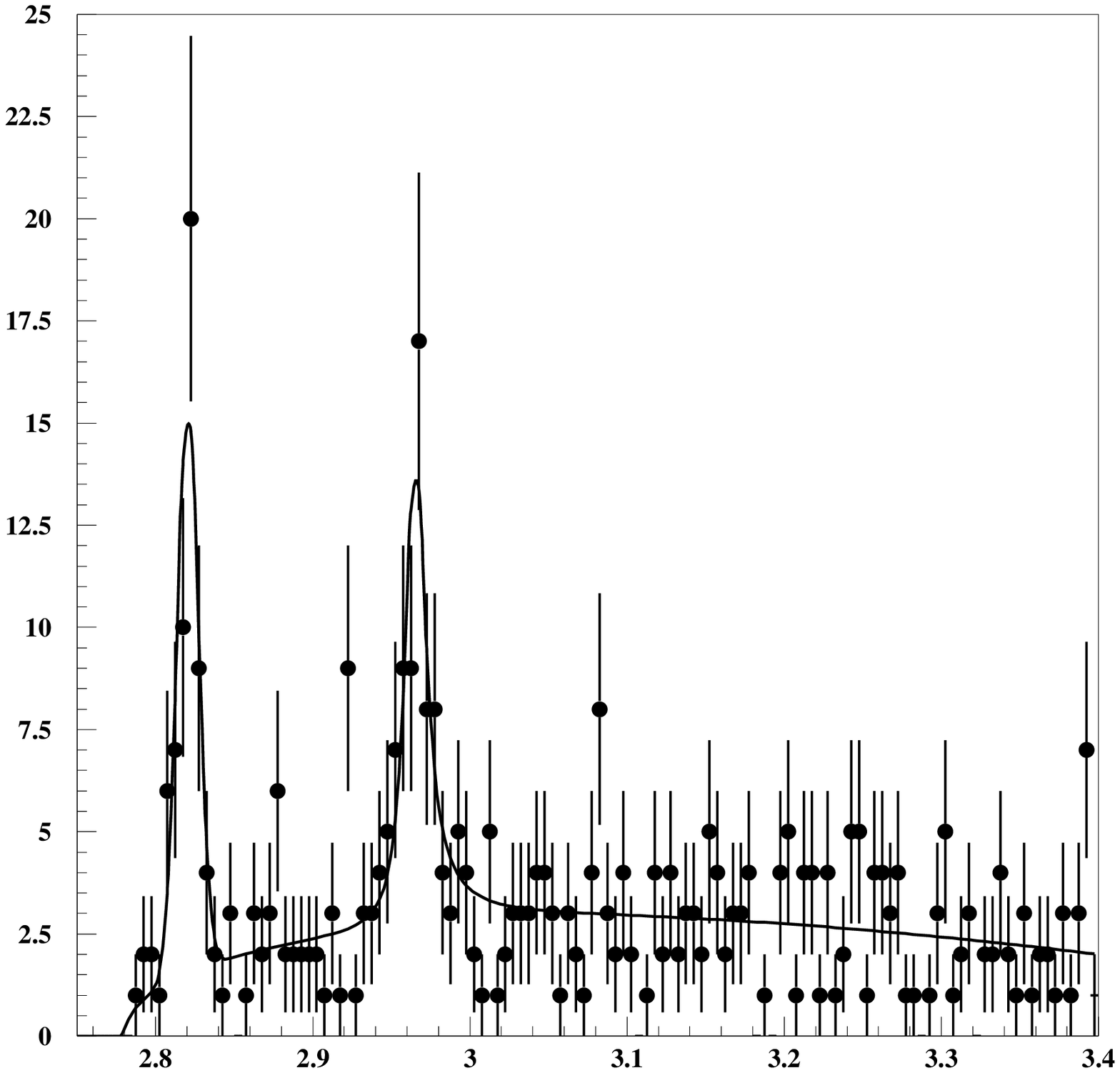}
    \end{picture}
  \end{minipage}
  \caption{
    Invariant mass distributions for (a) $\Xi_c(2645)^0 \pi^+$, and (b) $\Xi_c(2645)^+ \pi^-$.
    The $\Xi_c(2815)$ and $\Xi_c(2980)$ isodoublets are visible. From~\cite{bib:belleXc2980b}.
  }
  \label{fig:Xc2980}
\end{figure}

The $\Xi_c(2980)$ resonances are now well-established, having been seen with large significance
in more than one final state. In the discovery mode, $\Lambda_c^+ \overline{K} \pi$, the state
is close to the kinematic threshold of 2920~MeV$/c^2$---and as noted by BABAR, approximately half
of the $\Lambda_c^+ \overline{K} \pi$ decays proceed via an intermediate $\Sigma_c(2455)$,
for which the threshold is at 2950~MeV$/c^2$. The fit is therefore sensitive to the details of
the assumptions made about the signal lineshape, the fraction of decays proceeding via a
$\Sigma_c(2455)$, and the behaviour of the background near threshold; this led to mildly
inconsistent results for the mass and width from Belle and BABAR. The final state
$\Xi_c(2645) \pi$, reported by Belle and shown in Fig.~\ref{fig:Xc2980},
is well above threshold and does not suffer from these
complications; it also has a much lower background rate and does not require reconstruction
of a $K_S$ for either isospin partner. 

% Apologies for the ugly LaTeX; this is how the labels were put on the
% figures in the source paper.   -- Mat.

The narrow $\Xi_c(3077)$ resonances are also well-established: both isospin states have been seen
clearly by Belle and BABAR, and there is good agreement about the masses and widths.
BABAR has shown that most (if not all) of the decay proceeds via the quasi-two-body processes
$\Xi_c(3077) \to \Sigma_c(2455) \pi$ and $\Xi_c(3077) \to \Sigma_c(2520) \pi$ with roughly
equal branching fractions.

The $\Xi_c(3055)^+$ and $\Xi_c(3123)^+$ structures have been reported by BABAR in the
$\Lambda_c^+ K^- \pi^+$ final state only with statistical
significances of $6.4$ and $3.0$ standard deviations, respectively; they have not yet been
confirmed by Belle. The structures are not visible above background in the inclusive
$\Lambda_c^+ K^- \pi^-$ mass spectrum but emerge when requiring that the $\Lambda_c^+ \pi^-$
invariant mass lie in a narrow window around the $\Sigma_c(2455)^0$ (for the $\Xi_c(3055)^+$)
or the $\Sigma_c(2520)^0$ (for the $\Xi_c(3123)^+$).

BABAR has also searched the inclusive $\Lambda_c^+ \overline{K}$ and 
$\Lambda_c^+ \overline{K} \pi^+ \pi^-$ invariant mass spectra for evidence of further
narrow $\Xi_c$ states but did not find any significant structure above the background.

\begin{figure}[htb]
  \centering
  \includegraphics*[width=80mm]{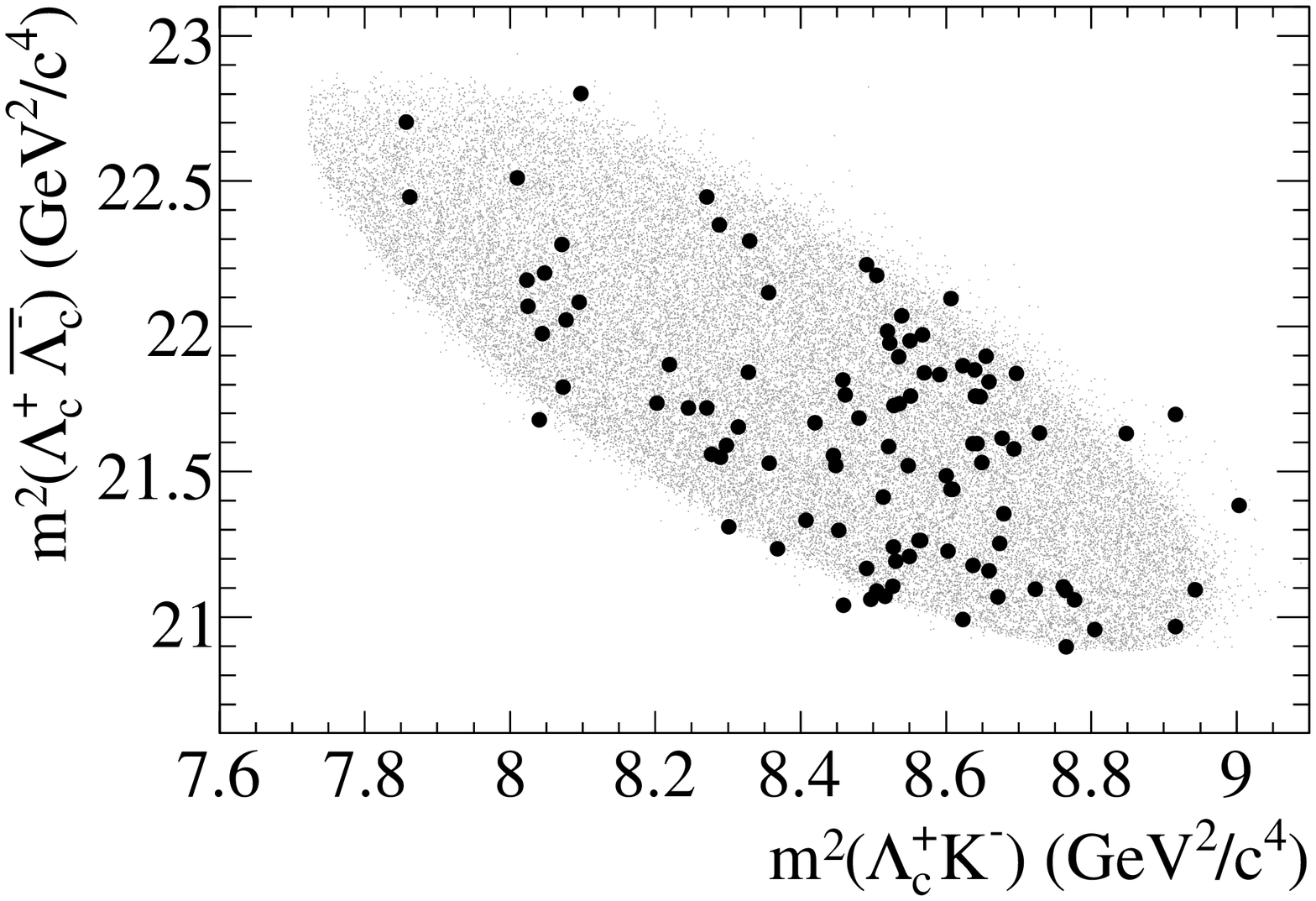}
  \includegraphics*[width=80mm]{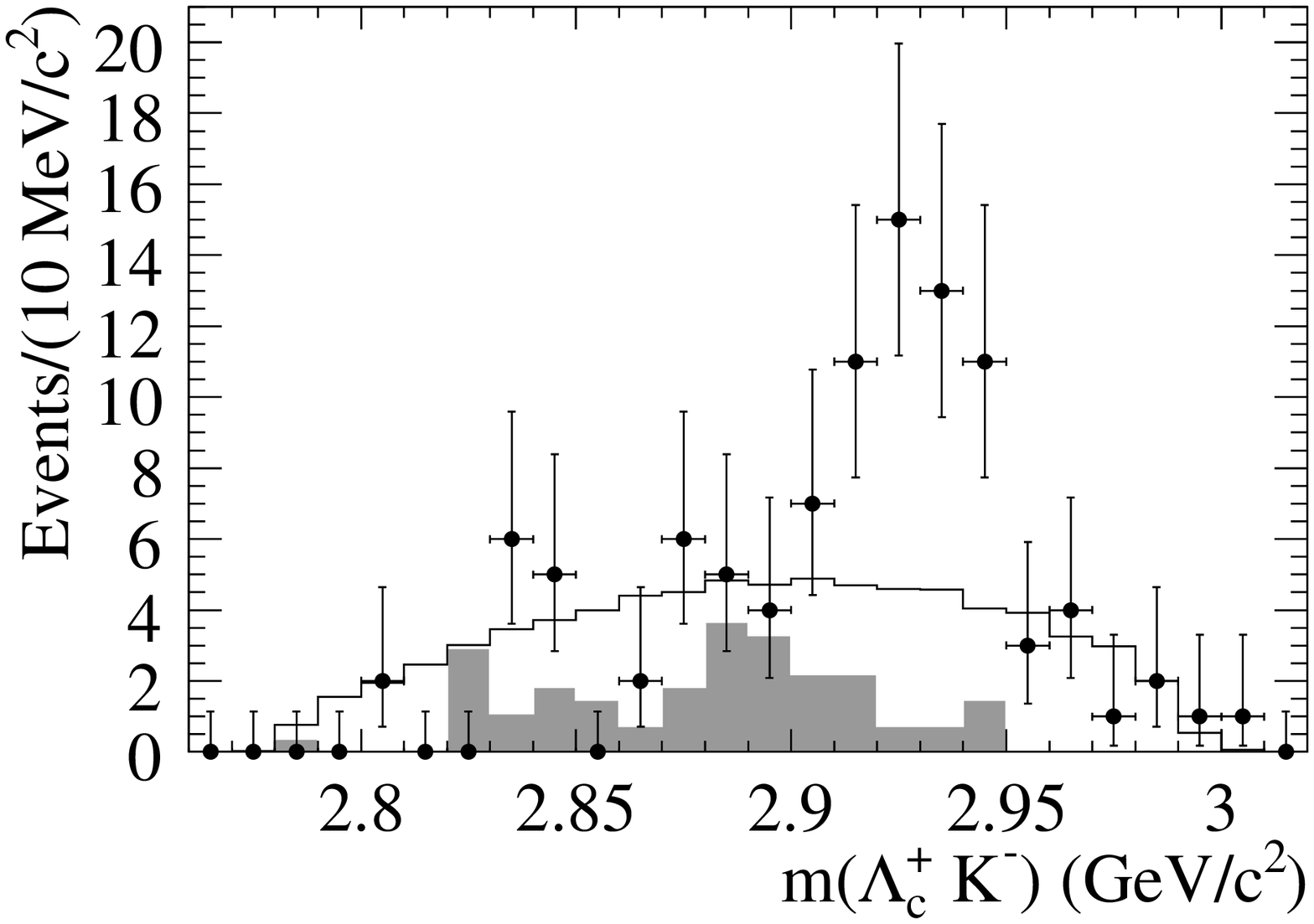}
  \caption{
    Reconstructed $B^- \to \Lambda_c^+ \bar{\Lambda}_c^- K^-$ candidates, showing
    a Dalitz plot (upper) and the $\Lambda_c^+ K^-$ invariant mass distribution (lower).
    Data from the signal region are shown as black points; signal from a
    phase-space simulation are shown as small grey points in~(a) and as
    a histogram in~(b); and data from a mass sideband region are shown as
    a shaded histogram in~(b), normalized to the expected background in the signal region.
    From~\cite{bib:babarB2LcLcK}.
  }
  \label{fig:Xc2930}
\end{figure}

There is one other candidate for a new $\Xi_c$ state, seen by BABAR in an analysis
of $B^- \to \Lambda_c^+ \bar{\Lambda}_c^- K^-$
in a sample of $230 \times 10^6$ $\Upsilon(4S) \to B \overline{B}$ decays~\cite{bib:babarB2LcLcK}. 
The Dalitz plot,
shown in Fig.~\ref{fig:Xc2930}, is clearly inconsistent with a phase-space
distribution and is consistent with a dominant contribution from a single
$\Xi_c^0$ resonance with mass $2931 \pm 3 \pm 5$~MeV$/c^2$ and
width $36 \pm 7 \pm 1$~MeV. However, this structure has not yet been
confirmed by Belle. The $B$ decay mode itself was first observed
by Belle in a sample of $386 \times 10^6$ $\Upsilon(4S) \to B \overline{B}$ decays~\cite{bib:belleB2LcLcK};
the publication included the $\Lambda_c^+ \bar{\Lambda}_c^-$
invariant mass spectrum as part of a search for charmonium states but not the full
Dalitz plot or the $\Lambda_c^+ K^-$ projection.

\subsection{$\Omega_c$ states}

\begin{figure}
  \centering
  \includegraphics*[width=80mm]{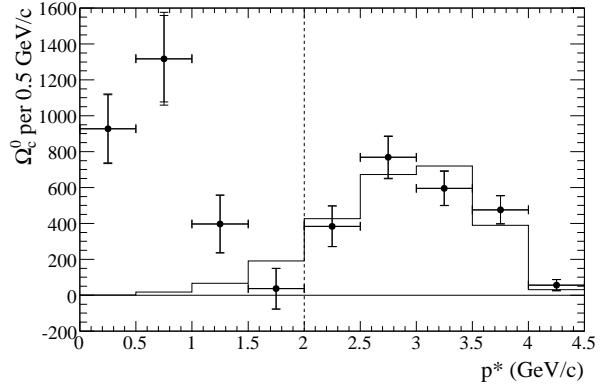}
  \caption{
    The inclusive momentum spectrum of $\Omega_c^0$ baryons
    in $e^+e^-$ data at $\sqrt{s}\approx10.6$~GeV$/c$.
    The data show two peaks: the one at higher momentum is
    due to $\Omega_c^0$ production in continuum events, and
    the one at lower momentum is due to the decays of $B$ mesons.
    The histogram shows the Bowler fragmentation function, fit
    to the data above 2~GeV$/c$. From~\cite{bib:babarOmegac}.
  }
\label{fig:OcPstar}
\end{figure}

The weakly-decaying $\Omega_c^0$ ground state has been established
for more than a decade~\cite{bib:pdg}, but until recently its properties
had only been measured with a handful of events. Both BABAR and Belle have
now published studies with large statistics, confirming and expanding upon
previous results. BABAR has reported the first observation of
$\Omega_c^0$ production in $B$ meson decays with an inclusive method,
as shown in Fig.~\ref{fig:OcPstar}, along with measurements of ratios of branching
fractions with much improved precision in 231~fb$^{-1}$ of data~\cite{bib:babarOmegac}. 
In addition, Belle has made a new measurement of the $\Omega_c^0$ mass
with 673~fb$^{-1}$ of data~\cite{bib:belleOmegac}, obtaining
$m(\Omega_c^0) = 2693.6 \pm 0.3 ^{+1.8}_{-1.5}$~MeV$/c^2$.

\begin{figure}
  \centering
  \includegraphics*[width=80mm,viewport=6 2 351 178,clip]{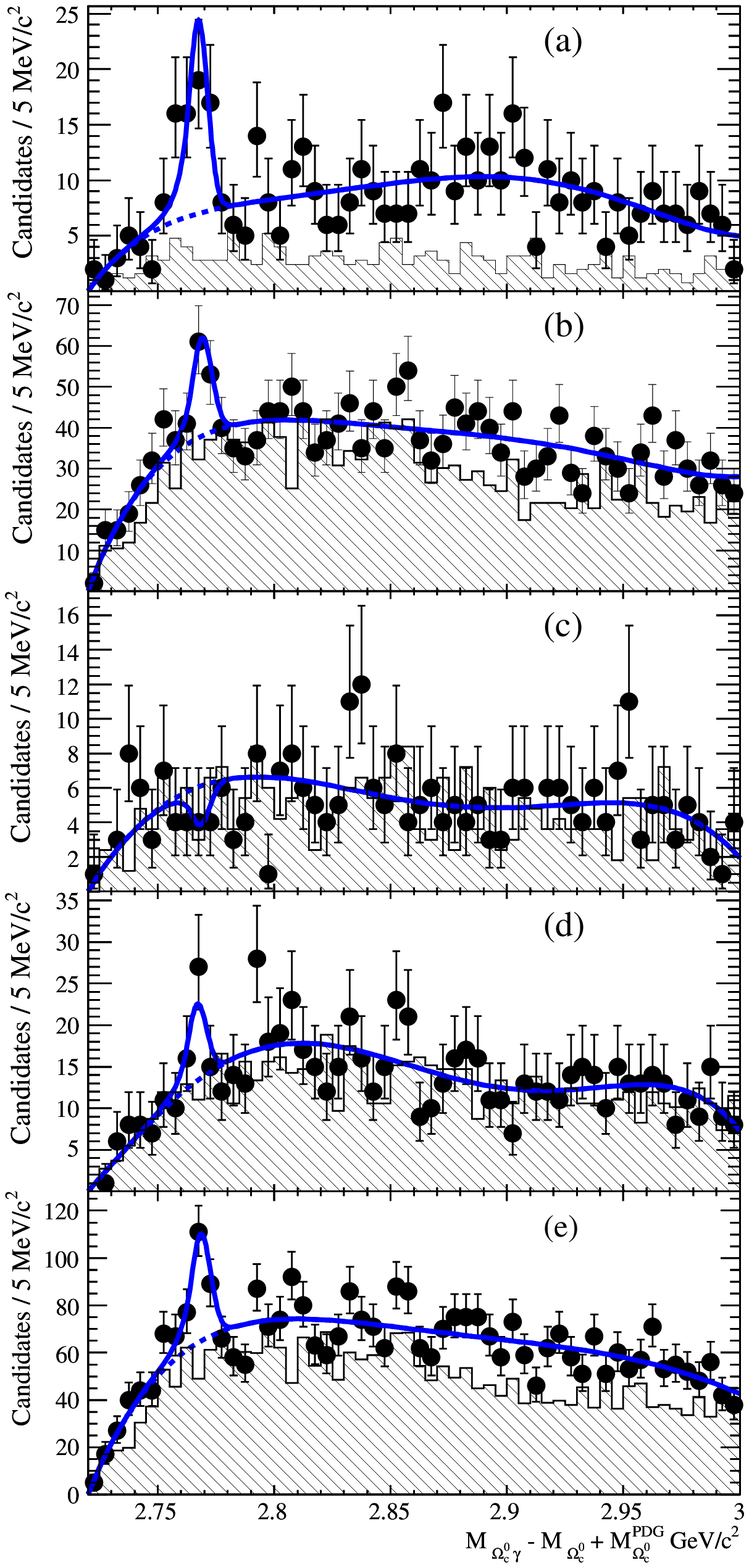}
  \includegraphics*[width=65mm]{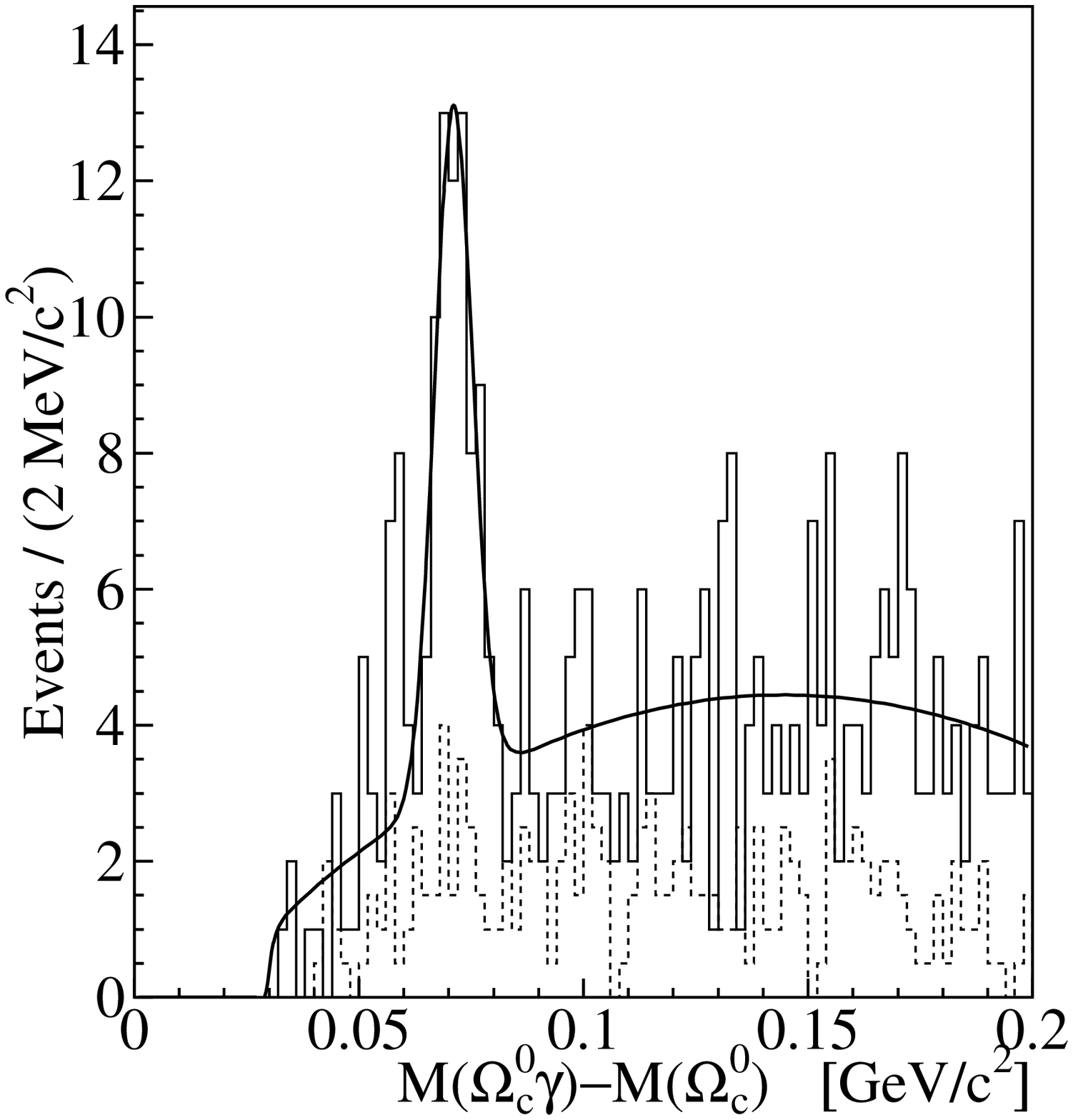}
  \caption{
    Invariant mass spectra of $\Omega_c \gamma$ at BABAR in 231~fb$^{-1}$ (upper)
    and Belle in 673~fb$^{-1}$ of data (lower). 
    The $\Omega_c(2770)^0$ resonance is visible in each plot.
    From~\cite{bib:babarOc2770} and~\cite{bib:belleOmegac}.
  }
\label{fig:Oc2770}
\end{figure}

The quark model predicts two $\Omega_c^0$ ground states:
a lower-mass octet state with $J^P = 1/2^+$ and a higher-mass
decuplet state with $J^P = 3/2^+$. This higher-mass partner was
observed by BABAR\cite{bib:babarOc2770} and confirmed by Belle~\cite{bib:belleOmegac}
(Fig.~\ref{fig:Oc2770}); this completes the experimental
discoveries of all $C=1$ ground states predicted by the quark model.
The mass splitting between the two $\Omega_c$ partner states $\Delta M$
is measured to be:
\begin{eqnarray*}
  \mbox{BABAR:} & \Delta M = & 70.8 \pm 1.0 \pm 1.1 ~ \mathrm{MeV}/c^2 \\
  \mbox{Belle:} & \Delta M = & 70.7 \pm 0.9 ^{+0.1}_{-0.9}  ~ \mathrm{MeV}/c^2 ,
\end{eqnarray*}
with excellent agreement between the two experiments.
Since $\Delta M$ is less than the pion mass, there is no allowed strong
decay for the heavier partner state and it therefore decays electromagnetically
to $\Omega_c^0 \gamma$.

From the constituent quark model~\cite{bib:quarkModelMass},
we would expect other octet-decuplet mass splittings,
$m(\Xi_c(2645))-m(\Xi_c^{\prime})$ and $m(\Sigma_c(2520))-m(\Sigma_c(2455))$,
to be comparable to the $\Omega_c$ splitting but with some variation
due to broken $SU(3)$ symmetry entering via the spin-spin coupling term. 
It is interesting just how small the variation is; averaging
over the isospin states and assuming the uncertainties
to be systematics-dominated, the current data give:~\cite{bib:pdg}
\begin{eqnarray*}
  m(\Xi_c(2645))-m(\Xi_c^{\prime}) &\approx& 69.5 \pm 3 ~ \mathrm{MeV}/c^2 \\
  m(\Sigma_c(2520))-m(\Sigma_c(2455)) &\approx& 64.3 \pm 0.5  ~ \mathrm{MeV}/c^2
  .
\end{eqnarray*}

\section{Production of charmed baryons}

The production of baryons in $e^+ e^-$ interactions and meson decay is
still a subject of active study. Charmed baryons provide an excellent
laboratory for probing baryon formation, especially at the $e^+ e^-$
$B$-factories where they are produced copiously.

\subsection{Two-body $B$ decays to baryons}

It has been known for many years that $B$ mesons decay to baryons
at a significant rate, with a branching fraction of approximately
7\%~\cite{bib:argusInclusiveBtoBaryons}.
However, individual decay modes---especially two-body and
quasi-two-body modes---have been challenging to observe.
The reason for this is two-fold. First, the branching
fractions are often (but not always) small, e.g.
  $\mathcal{B}(\overline{B}^0 \to \Lambda_c^+ \bar{p}) = (2.0 \pm 0.4) \times 10^{-5}$,
and
  $\mathcal{B}(\overline{B}^0 \to p \bar{p}) < 1.1 \times 10^{-7}$
at the 90\% confidence limit~\cite{bib:pdg}.
Second, non-CKM-suppressed decays involve a charmed particle which
must be reconstructed, implying another factor of $\mathcal{O}(\mathrm{few}\%)$
in the yield due to the charm branching fraction.

One notable exception to the first point is $B^- \to \Xi_c^0 \bar{\Lambda}_c^-$, which was
observed by Belle in an analysis of $386 \times 10^6$ $\Upsilon(4S) \to B \overline{B}$ 
decays~\cite{bib:belleB2XcLc} (along with evidence for its isospin partner mode
$B^- \to \Xi_c^0 \bar{\Lambda}_c^-$) and confirmed by BABAR
with a sample of $230 \times 10^6$ $\Upsilon(4S) \to B \overline{B}$ decays~\cite{bib:babarB2LcLcK}.
They obtained the following values for
$\mathcal{B}(B^- \to \Xi_c^0 \bar{\Lambda}_c^-) \times
\mathcal{B}(\Xi_c^0 \to \Xi^- \pi^+)$:
\begin{eqnarray*}
  \mbox{Belle:} & (4.8 ^{+1.0}_{-0.9} \pm 0.9 \pm 1.0) \times 10^{-5} \\
  \mbox{BABAR:} & (2.1 \pm 0.7 \pm 0.3 \pm 0.5) \times 10^{-5} 
  .
\end{eqnarray*}
The value of $\mathcal{B}(\Xi_c^0 \to \Xi^- \pi^+)$ has not been measured
experimentally but is expected to be of order 1--2\%~\cite{bib:theoryXicBF},
which would imply $\mathcal{B}(B^- \to \Xi_c^0 \bar{\Lambda}_c^-) \sim 1 \times 10^{-3}$.
This is large compared to other two-body baryonic decay modes, although
still smaller than corresponding mesonic decay modes
(e.g. $\mathcal{B}(B^- \to D_s^- D^0) = (1.0 \pm 0.2)\%$~\cite{bib:pdg}).

\subsection{Threshold enhancements in $B$ decays to baryons}

One explanation for this pattern
is that producing a baryon-antibaryon pair with large energy release 
(i.e. where the baryon and antibaryon have large relative velocity)
is kinematically suppressed, since such a reaction requires two
hard gluons~\cite{bib:theoryHouSoni,bib:theoryCheng}.
This turns the usual intuition on its head: processes with a larger
phase space have smaller rates. A corollary is that for three-body
(or, in general, multi-body) baryonic $B$ decays with a large energy
release, the bulk of the Dalitz plane is kinematically suppressed---but
not the region close to the baryon-antibaryon threshold. This implies
a kinematic enhancement near the threshold.

\begin{figure}
  \centering
  \includegraphics*[width=70mm,viewport=0 8 553 430,clip]{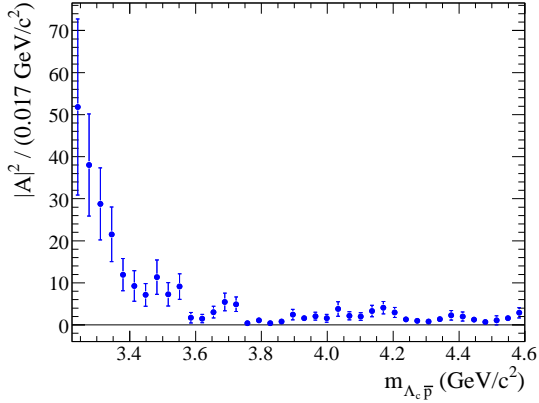}
  \includegraphics*[width=60mm,viewport=0 8 372 377,clip]{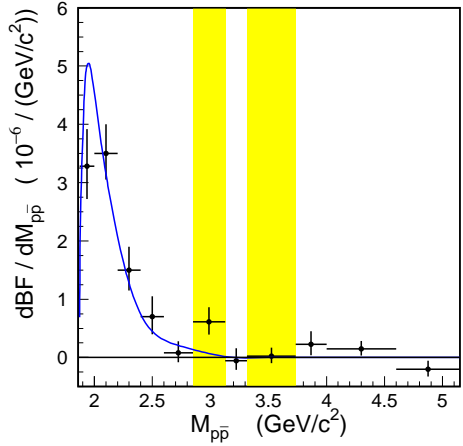}
  \includegraphics*[width=60mm,viewport=0 8 372 377,clip]{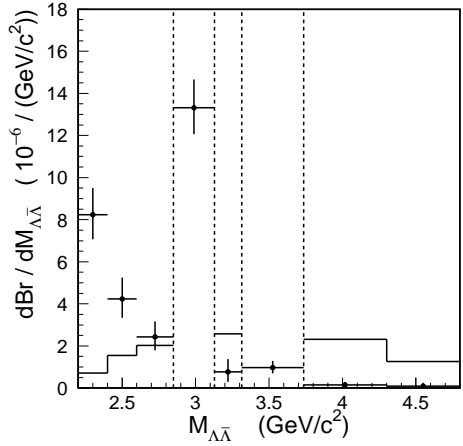}
  \caption{
    Threshold enhancements.
    The upper plot (from~\cite{bib:babarStephanie})
    shows the projected amplitude squared for
    $B^- \to \Lambda_c^+ \bar{p} \pi^-$ decays near
    threshold after efficiency correction, background
    subtraction, and correction for the three-body
    phase space.
    The middle plot (from~\cite{bib:belleB2ppK})
    shows the differential branching fraction
    for $B^+ \to p \bar{p} \pi^+$; the shaded/yellow
    bands indicate the two principal charmonium regions
    and the solid curve is a theoretical prediction~\cite{bib:theoryCheng}.
    The lower plot (from~\cite{bib:belleB2LLK})
    shows the differential branching fraction
    for $B^- \to \Lambda \bar{\Lambda} K^-$; the dashed
    lines indicate the two principal charmonium regions,
    and the solid histogram is a phase-space simulation.
    The analyses were performed on samples of
    383,
    449, and
    $657 \times 10^6$ $\Upsilon(4S) \to B \overline{B}$ decays,
    respectively.
  }
  \label{fig:thresholdPeaks}
\end{figure}

\begin{figure}
\centering
  \includegraphics*[width=80mm]{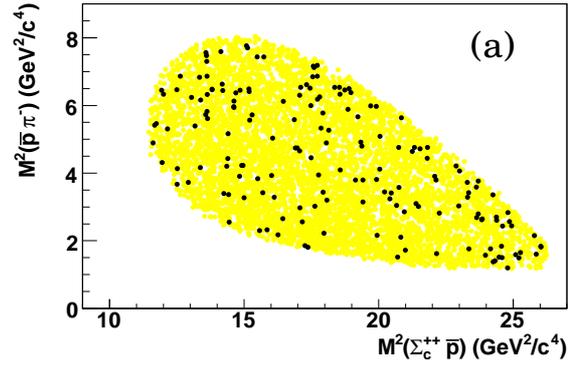}
  \includegraphics*[width=80mm]{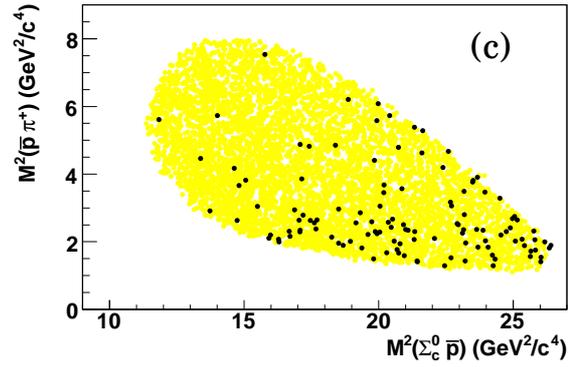}
\caption{
  Dalitz plots for 
  $\overline{B}^0 \to \Sigma_c(2455)^{++} \bar{p} \pi^{-}$ (upper) and
  $\overline{B}^0 \to \Sigma_c(2455)^{0}  \bar{p} \pi^{+}$ (lower).
  The black points represent the data and the
  yellow/light points are simulated according to three-body phase space.
  A sample of $388 \times 10^6$ $\Upsilon(4S) \to B \overline{B}$ decays
  were used. From~\cite{bib:belleB2Scppi}.
}
\label{fig:B2Scppi}
\end{figure}

Threshold peaks have been seen in a number of baryonic $B$ decays
(some illustrated in Fig.~\ref{fig:thresholdPeaks}) including
  $B^- \to \Lambda_c^+ \bar{p} \pi^-$,
  $B^- \to p \bar{p} \pi^-$,
  $B^- \to p \bar{p} K^-$,
  $B^- \to \Lambda \bar{\Lambda} K^-$,
  $B^- \to \Lambda \bar{p} \gamma$, and
  $B^0 \to p \bar{\Lambda} \pi^-$
\cite{bib:babarStephanie,bib:belleB2ppK,bib:belleB2pLg,bib:belleB2LLK},
as well as in other decays such as $J/\psi \to p \bar{\Lambda} K^-$
and $J/\psi \to p \bar{p} \gamma$
\cite{bib:besJpsi2pLK,bib:besJpsi2ppg}, so there does appear to be a general pattern. 
However, there are other mechanisms which can also cause a threshold peak---most
obviously a resonance at or just below the threshold, but also other
effects such as final-state interactions~\cite{bib:theoryBugg,bib:theoryKerbikov,bib:theoryChen}---so 
the nature of the structure in a particular mode may be more complicated than this.
There are also modes where no threshold peak is seen even though it
might be expected, e.g.
  $\overline{B}^0 \to \Sigma_c(2455)^{++} \bar{p} \pi^-$ and
  $\overline{B}^0 \to \Sigma_c(2455)^{0}  \bar{p} \pi^+$~\cite{bib:belleB2Scppi}
and
  $J/\psi \to p \bar{p} \pi^0$~\cite{bib:besJpsi2ppg}.
The former is illustrated in Fig.~\ref{fig:B2Scppi}; the Dalitz plot
for $\overline{B}^0 \to \Sigma_c(2455)^{0} \bar{p} \pi^{+}$ in
particular is strikingly empty close to the baryon-antibaryon threshold.

\subsection{Baryon formation in the continuum}

\begin{figure}
\centering
\includegraphics*[width=80mm,viewport=28 29 419 423,clip]{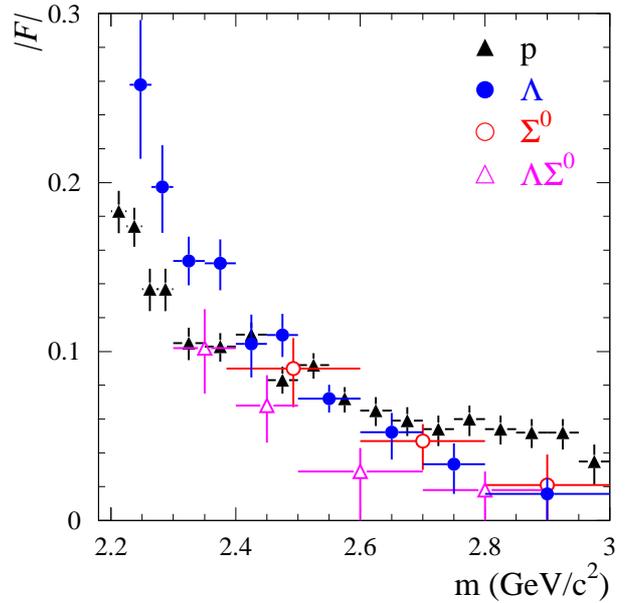}
\caption{
  The effective baryon form factors as a function of baryon-antibaryon
  invariant mass, measured with ISR events in a sample of
  230~fb$^{-1}$ of $e^+ e^-$ data.
  From~\cite{bib:babarISRhyperon}.
}
\label{fig:hyperonISR}
\end{figure}

Enhanced rates near a baryon-antibaryon threshold have also been seen in
$e^+ e^-$ events with an initial-state radiation (ISR) photon. BABAR has
measured the effective form factors for 
  $p\bar{p}$, 
  $\Lambda \bar{\Lambda}$,
  $\Sigma^0 \overline{\Sigma}^0$, and
  $\Lambda \overline{\Sigma}^0$
close to threshold~\cite{bib:babarISRpp,bib:babarISRhyperon};
the distributions are shown in Fig.~\ref{fig:hyperonISR}.
In each case we see a strong enhancement at or close to threshold.

\begin{figure}
\centering
  \includegraphics*[width=80mm]{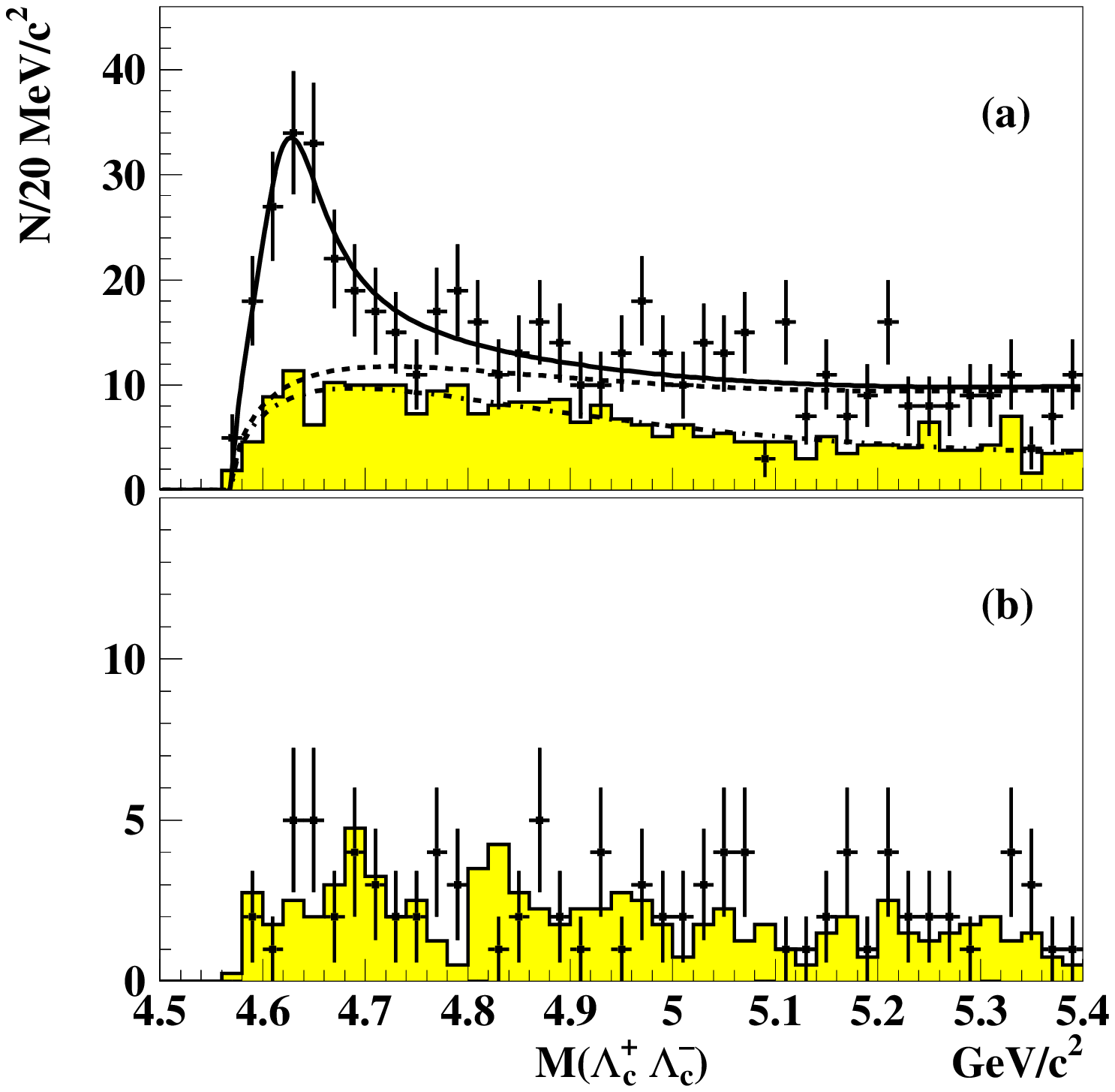}
  \includegraphics*[width=80mm]{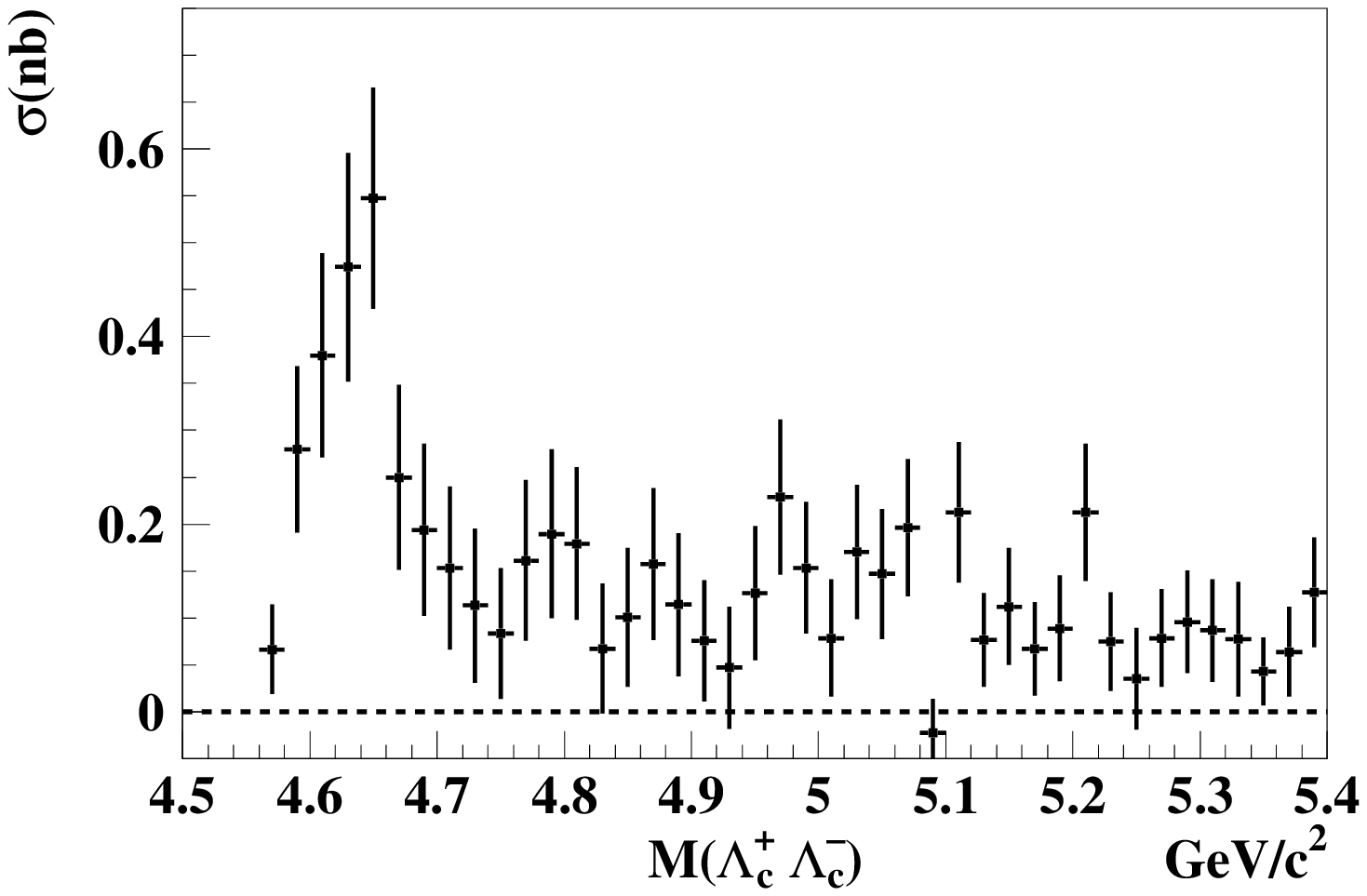}
\caption{
  The exclusive process
  $e^+ e^- \to \Lambda_c^+ \bar{\Lambda}_c^-$,
  measured in $e^+ e^- \to \Lambda_c^+ \bar{\Lambda}_c^- \gamma_{\mathrm{ISR}}$ events.
  Plot (a) shows the $\Lambda_c^+ \bar{\Lambda}_c^-$ invariant mass
  spectrum for events with a $\bar{p}$ tag that lie in the signal region (points)
  or $\Lambda_c^+$ sidebands (shaded histogram). The dot-dashed, dashed, and solid
  curves represent combinatoric background, a threshold function, and an $S$-wave relativistic
  Breit-Wigner function cumulatively. 
  Plot (b) shows equivalent events from the background region.
  The lower plot shows the corresponding cross-section for
  $e^+ e^- \to \Lambda_c^+ \bar{\Lambda}_c^-$ (after correcting for
  background, efficiency, and the differential ISR luminosity).  
  Errors shown are statistical only.
  695~fb$^{-1}$ of data were used.
  From~\cite{bib:belleLcLcISR}.
}
\label{fig:LcLcISR}
\end{figure}

In the charm sector, Belle has
recently reported a large structure in
$e^+ e^- \to \Lambda_c^+ \bar{\Lambda}_c^- \gamma_{\mathrm{ISR}}$
events, shown in Fig.~\ref{fig:LcLcISR}~\cite{bib:belleLcLcISR}.
They note that the nature of the enhancement is unclear, but that
it can be described by an $S$-wave relativistic Breit-Wigner with
mass $4634 ^{+8}_{-7}$$^{+5}_{-8}$~MeV$/c^2$ and
width $92 ^{+40}_{-24}$$^{+10}_{-21}$~MeV.
It seems quite plausible that this is one of the many
charmonium and charmonium-like states above the
$D \overline{D}$ threshold~\cite{bib:pdg}.
In particular, it may be the $Y(4660)$ seen previously in
$e^+ e^- \to (J/\psi \pi^+ \pi^-) \gamma_{\mathrm{ISR}}$ events~\cite{bib:belleJpsipipi}
with mass $4664 \pm 11 \pm 5$~MeV$/c^2$ and
width $48 \pm 15 \pm 3$~MeV; the modest discrepancy in parameters
between the two measurements may be due to differing form factors
for the two final states~\cite{bib:theoryBugg4660}.

The kinematic favourability of forming baryon-antibaryon pairs
with small relative velocity is certainly not a new discovery
or unique to low-energy interactions; it was noted at LEP-I,
for example, that correlated $p \bar{p}$, $\Lambda \bar{\Lambda}$,
and $\Lambda \bar{p}$ pairs peak at small values of the
rapidity difference $|\Delta y|$~\cite{bib:delphi,bib:opal}
and are predominantly produced in the same jet.

\begin{figure}
  \centering
  \includegraphics*[width=80mm]{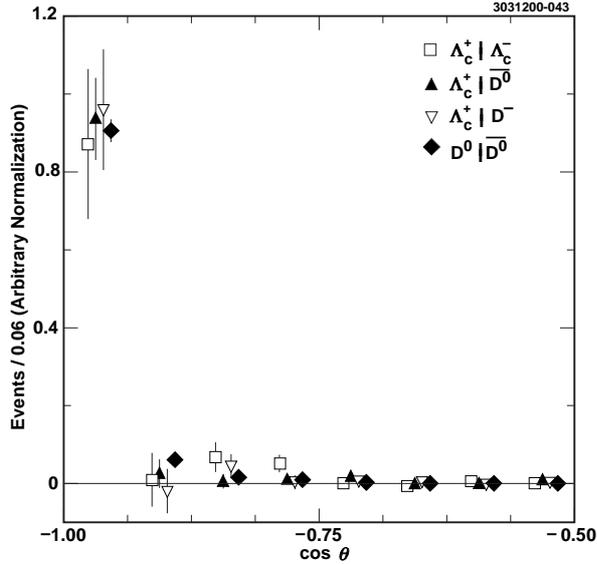}
  \caption{
    The production angle distribution between various combinations of
    charmed baryons and charmed mesons.
    In each case, the charmed particles are approximately
    back-to-back.
    From~\cite{bib:cleoLcLc}.
  }
  \label{fig:cleoLcLc}
\end{figure}

This is not universally true, however. In 13.6~fb$^{-1}$ of $e^+ e^-$
data at $\sqrt{s} \sim 10.6$~GeV, CLEO studied a sample of
continuum events containing both a $\Lambda_c^+$ and a $\bar{\Lambda}_c^-$~\cite{bib:cleoLcLc}.
(A momentum cut on the $\Lambda_c^+$ was used to exclude those
baryons produced in $B$ meson decays.)
As illustrated in Fig.~\ref{fig:cleoLcLc}, the charmed baryons
in these events are strongly peaked towards $\cos \theta = -1$,
i.e. with the baryons in opposite hemispheres. Broadly, there
are three possible classes of such events:
\begin{enumerate}
  \item Events in which the two charm jets fragment independently
    with one producing a $\Lambda_c^+$ and a light antibaryon and
    the other producing a $\bar{\Lambda}_c^-$ and a light baryon.
    Baryon number is conserved in each jet, and the invariant mass
    of and rapidity gap between each baryon-antibaryon pair are small.
  \item As (1), except that the fragmentation of the two jets is
    correlated such that producing a $\Lambda_c^+$ on one side
    makes it more likely that a $\bar{\Lambda}_c^-$ is produced
    on the other side.
    Again, baryons are produced in pairs which have invariant
    mass close to theshold, and baryon number is conserved locally.
  \item Events with primary baryon production that contain
    no baryons besides the $\Lambda_c^+$ and $\Lambda_c^-$.
    Baryon number is not conserved in each jet and the
    baryon-antibaryon pair is far from threshold.
\end{enumerate}
CLEO ruled out the first class as the sole source of the events observed, since
the rate was approximately 3.5 times higher than would be expected given independent
fragmentation.
Qualitatively, they looked for additional protons in the event and did
not find signal; this disfavours the second class (since those events have
two additional light baryons) but does not exclude it.

\begin{figure}
  \centering
  \includegraphics*[width=80mm]{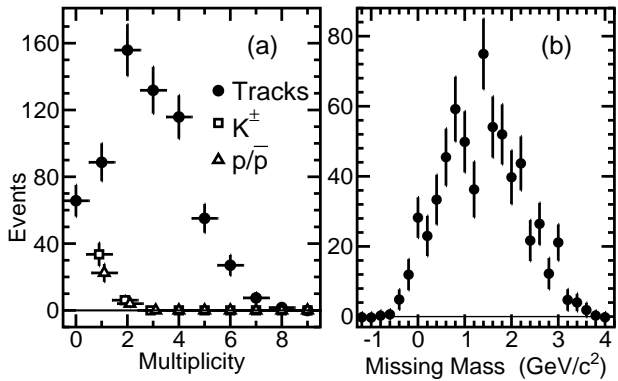}
  \caption{
    Background-subtracted distributions for events containing
    $\Lambda_c^+ \bar{\Lambda}_c^-$: (a) the number of
    additional tracks, identified $K^{\pm}$, and
    additional identified $p/\bar{p}$; and (b) the missing mass,
    with imaginary masses given negative real values.
    Most events have no identified $K^{\pm}$
    or $p/\bar{p}$ amd the corresponding
    zero-multiplicity points are off the vertical
    scale in~(a).
    Preliminary BABAR result---see~\cite{bib:babarLcLc}.
  }
  \label{fig:babarLcLc}
\end{figure}

BABAR has made a preliminary study of this process with 220~fb$^{-1}$ of data~\cite{bib:babarLcLc}.
They see an excess of $\Lambda_c^+ \bar{\Lambda}_c^-$ events by a factor of 4, confirming the CLEO
observation. In addition, they find that most of the events seen have zero additional
identified protons, such that the rate seen is consistent with that expected from
material interactions and from misidentification of other charged particles (Fig.~\ref{fig:babarLcLc}(a)).
Further, when plotting the missing mass distribution of the event (after subtracting the momenta of
the $\Lambda_c^+$, $\bar{\Lambda}_c^-$, and any reconstructed tracks), they find that most events
have well below 2~GeV$/c^2$ of missing mass, and are therefore inconsistent with containing
a baryon and an antibaryon that are not reconstructed or are misidentified as pions (Fig.~\ref{fig:babarLcLc}(b)).
While some contribution from classes (1) and (2) is not ruled out, this implies that
class (3) is the dominant effect. 

\section{Conclusions}

There has been a great deal of progress in charmed baryon spectroscopy recently, and
the high statistics of Belle and BABAR are still uncovering new states. The use of
$B$ meson decays to study resonant states in a fully exclusive environment is a
particularly powerful tool; hopefully it will be extended beyond the
$\Sigma_c(2455)$ to measure the quantum numbers of other baryon states.

There are also a number of results finding evidence for threshold enhancements in
multi-body baryonic decays. The overall pattern is not clear here, particularly
given the absence of an enhancement in some modes. We look
forward to further experimental data and to input from phenomenology to
help understand these structures.


\begin{thebibliography}{99}

\bibitem{bib:charm2009}
M. J. Charles, ``Charmed Baryons'',
presented at CHARM~2009.

\bibitem{bib:disclaimer2}
Charge-conjugate processes are implied throughout.

\bibitem{bib:disclaimer1}
Unless otherwise stated, the first quoted uncertainty is statistical,
the second is systematic, and the third (if present) is due to uncertainty
in the $\Lambda_c^+ \to p K^- \pi^+$ branching fraction.

\bibitem{bib:babarDetector}
  BABAR Collaboration, 
  B. Aubert et al.,
  Nucl. Instrum. Meth. \textbf{A479} (2002) 1.

\bibitem{bib:belleDetector}
  Belle Collaboration,
  A. Abashian et al.,
  Nucl. Instrum. Meth. \textbf{A479} (2002) 117.

\bibitem{bib:pdg}
  Particle Data Group,
  C. Amsler et al.,
  Physics Letters \textbf{B667} (2008) 1.

\bibitem{bib:babarStephanie}
  BABAR Collaboration,
  B. Aubert et al.,
  Phys. Rev. \textbf{D78} (2008) 112003.

\bibitem{bib:belleStephanie}
  Belle Collaboration,
  N. Gabyshev et al.,
  Phys. Rev. Lett. \textbf{97} (2006) 242001.

\bibitem{bib:belleSc2800}
  Belle Collaboration,
  R. Mizuk et al.,
  Phys. Rev. Lett. \textbf{94} (2005) 122002.

\bibitem{bib:belleXc2980}
  Belle Collaboration,
  R. Chistov et al.,
  Phys. Rev. Lett. \textbf{97} (2006) 162001.

\bibitem{bib:babarXc2980}
  BABAR Collaboration,
  B. Aubert et al.,
  Phys. Rev. \textbf{D77} (2008) 012002.

\bibitem{bib:belleXc2980b}
  Belle Collaboration,
  T. Lesiak et al.,
  Phys. Lett. B \textbf{665} (2008) 9.

\bibitem{bib:babarB2LcLcK}
  BABAR Collaboration,
  B. Aubert et al.,
  Phys. Rev. \textbf{D77} (2008) 031101.

\bibitem{bib:belleB2LcLcK}
  Belle Collaboration,
  N. Gabyshev et al.,
  Phys. Rev. Lett. \textbf{97} (2006) 202003.

\bibitem{bib:babarOmegac}
  BABAR Collaboration,
  B. Aubert et al.,
  Phys. Rev. Lett. \textbf{99} (2007) 062001.

\bibitem{bib:belleOmegac}
  Belle Collaboration,
  E. Solovieva et al.,
  Phys. Lett. B \textbf{672} (2009) 1.

\bibitem{bib:babarOc2770}
  BABAR Collaboration,
  B. Aubert et al.,
  Phys. Rev. Lett. \textbf{97} (2006) 232001.

\bibitem{bib:quarkModelMass}
  See, e.g., 
  S.~Gasiorowicz and J.L.~Rosner,
  Am.\ J.\ Phys.\  {\bf 49} (1981) 954.

\bibitem{bib:argusInclusiveBtoBaryons}
  ARGUS Collaboration,
  H.~Albrecht et al.,
  Z.\ Phys.\  C {\bf 56} (1992) 1.

\bibitem{bib:belleB2XcLc}
  Belle Collaboration,
  R. Chistov et al.,
  Phys. Rev. \textbf{D74} (2006) 111105.

  %``Quark And Pole Models Of Nonleptonic Decays Of Charmed Baryons,''
\bibitem{bib:theoryXicBF}
  P.~Zenczykowski,
  Phys.\ Rev.\  D {\bf 50} (1994) 402.

\bibitem{bib:theoryHouSoni}
  W.S.~Hou and A.~Soni,
  Phys.\ Rev.\ Lett.\  {\bf 86} (2001) 4247.

\bibitem{bib:theoryCheng}
  H.Y.~Cheng,
  Int.\ J.\ Mod.\ Phys.\  A {\bf 21} (2006) 4209.

% ``Study of the decay mechanism for B+ to p pbar K+ and B+ to p pbar pi+,''
\bibitem{bib:belleB2ppK}
  Belle Collaboration,
  J.T.~Wei et al.,
  Phys.\ Lett.\  B {\bf 659} (2008) 80.

  %``Study of ${B^{+}} \to {p\bar{\Lambda}\gamma}$, ${p\bar{\Lambda}\pi^0}$ and
  %${B^{0}} \to {p\bar{\Lambda}\pi^-}$,''
\bibitem{bib:belleB2pLg}
  Belle Collaboration,
  M.Z.~Wang et al.,
  Phys.\ Rev.\  D {\bf 76} (2007) 052004.

  %``Observation of B0 to Lambda Lambdabar K0 and B0 to Lambda Lambdabar K*0 at
  %Belle,''
\bibitem{bib:belleB2LLK}
  Belle Collaboration,
  Y.W.~Chang et al.,
  Phys.\ Rev.\  D {\bf 79} (2009) 052006.

  %``Observation of a threshold enhancement in the p \bar{Lambda} invariant mass
  %spectrum,''
\bibitem{bib:besJpsi2pLK}
  BES Collaboration,
  M.~Ablikim et al.,
  Phys.\ Rev.\ Lett.\  {\bf 93} (2004) 112002.

  %``Observation of a near-threshold enhancement in th p pbar mass spectrum from
  %radiative J/psi-->gamma p pbar decays,''
\bibitem{bib:besJpsi2ppg}
  BES Collaboration,
  J.Z.~Bai et al.,
  Phys.\ Rev.\ Lett.\  {\bf 91} (2003) 022001.

  %``Reinterpreting several narrow `resonances' as threshold cusps,''
\bibitem{bib:theoryBugg}
  D.V.~Bugg,
  Phys.\ Lett.\  B {\bf 598} (2004) 8.

  %``Model-independent view on the low-mass proton-antiproton enhancement,''
\bibitem{bib:theoryKerbikov}
  B.~Kerbikov, A.~Stavinsky and V.~Fedotov,
  Phys.\ Rev.\  C {\bf 69} (2004) 055205.

  %``Rescattering Effect and Near Threshold Enhancement of $p\bar p$ System,''
\bibitem{bib:theoryChen}
  G.Y.~Chen, H.R.~Dong and J.P.~Ma,
  Phys.\ Rev.\  D {\bf 78} (2008) 054022.

  %``Study of intermediate two-body decays in $\bar{B}^0\to
  %\Sigma_c(2455)^{0}\bar{p}\pi^{+}$,''
\bibitem{bib:belleB2Scppi}
  Belle collaboration,
  H.O.~Kim et al.,
  Phys.\ Lett.\  B {\bf 669} (2008) 287.

  %``A Study of $e^{+} e^{-} \to p \bar{p}$ using initial state radiation with
  %BABAR,''
\bibitem{bib:babarISRpp}
  BABAR Collaboration,
  B.~Aubert et al.,
  Phys.\ Rev.\  D {\bf 73} (2006) 012005.

  %``Study of $e^{+} e^{-} \to \Lambda \bar{\Lambda}$, $\Lambda \bar{\Sigma}^0$,
  %$\Sigma^0 \bar{\Sigma}^0$ using initial state radiation with BABAR,''
\bibitem{bib:babarISRhyperon}
  BABAR Collaboration,
  B.~Aubert et al.,
  Phys.\ Rev.\  D {\bf 76} (2007) 092006.

  %``Observation of a near-threshold enhancement in the $e^+e^- \to \Lambda_c^+
  %\Lambda_c^-$ cross section using initial-state radiation,''
\bibitem{bib:belleLcLcISR}
  Belle Collaboration,
  G.~Pakhlova et al.,
  Phys.\ Rev.\ Lett.\  {\bf 101} (2008) 172001.

\bibitem{bib:belleJpsipipi}
  Belle Collaboration,
  X.L.~Wang et al.,
  Phys.\ Rev.\ Lett.\  {\bf 99} (2007) 142001.

  %``An alternative fit to Belle mass spectra for DD, D*D* and Lambda_C
  %Lambda_c,''
\bibitem{bib:theoryBugg4660}
  D.V.~Bugg,
  J.\ Phys.\ G {\bf 36} (2009) 075002.

\bibitem{bib:delphi}
  DELPHI Collaboration,
  P. Abreu et al.,
  Phys.\ Lett.\  B {\bf 416} (1998) 247.

  %``A study of parton fragmentation in hadronic Z0 decays using Lambda
  %Antilambda correlations,''
\bibitem{bib:opal}
  OPAL Collaboration,
  G.~Abbiendi et al.,
  Eur.\ Phys.\ J.\  C {\bf 13} (2000) 185.

  %``Correlated Lambda/c+ anti-Lambda/c- production in e+ e- annihilations  at
  %s**(1/2) approx. 10.5-GeV,''
\bibitem{bib:cleoLcLc}
  CLEO Collaboration,
  A.~Bornheim et al.,
  Phys.\ Rev.\  D {\bf 63} (2001) 112003.

\bibitem{bib:babarLcLc}
  B.L.~Hartfiel,
  Ph.D. thesis, UCLA, 2005;
  SLAC-R-823.

\end{thebibliography}
\end{document}